\documentclass[]{aa}
\usepackage{graphicx}
\usepackage[]{natbib}
\bibpunct{(}{)}{;}{a}{}{,}
\def\simlt{\lower.5ex\hbox{$\; \buildrel < \over \sim \;$}}
\def\simgt{\lower.5ex\hbox{$\; \buildrel > \over \sim \;$}}

\begin{document}

   \msnr{Df053}

   \title{The Stellar Winds of Galactic Centre and the Low Accretion Rate
          of Sgr~A*}

   \author{R.F. Coker}

   \institute{Department of Physics and Astronomy, \\ University of Leeds, 
              Leeds LS2 9JT  UK \\
              email: robc@ast.leeds.ac.uk
             }

   \date{Received 5 June 2001 / Accepted 18 June 2001}

   \titlerunning{Winds of Galactic Centre \& Sgr~A*}
   \authorrunning{Coker}

   \abstract{
An attempt is made to reconcile the large wind-loss rates of stars in
the Galactic Centre (GC) with the predicted low accretion rate 
for Sgr~A*, the putative blackhole at the heart of the Milky
Way.  It is found that, independent of the details of the accretion,
the bound but unaccreted gas has been accumulating in the potential well
of Sgr~A* for $\simlt 10^3$ yrs and thus is not in equilibrium.
Otherwise, the gas flows of the region would be visible in both the IR
and X-ray.  It appears that the blackhole was more active in the recent
past due to the passing of a supernova blast shock but is presently
in a short-lived dormant phase.  The extended low frequency radio emission
from the central parsec should visibily increase over the next few decades,
as the shock passes completely in front of the absorbing gas and dust
near Sgr~A*.  The GC may become more active in
$\simlt 10^5$ yrs due to either another supernova or sufficient accumulation of
stellar winds in the central arcsecond.
   }

   \maketitle

\section{Introduction}

Sgr~A*, the stationary, compact, nonthermal radio source
at the Galactic Centre (GC), appears to be coincident
with a $2.6\pm0.2\times10^6 M_{\sun}$ point-like object
\citep{EG96, EG97, GKMB98, BS99, GPEGO00, GMBTK00}.
It is probable that Sgr~A* is associated with the accretion
of matter onto a supermassive blackhole
\citep[for a recent review of the evidence, see][]{GE99}.
Models for the accretion process itself vary widely,
including simplistic spherical accretion \citep{CM00,CM01},
an advective flow \citep{NYM95}, a convective flow \citep{SPB99},
a compact jet \citep{FM00}, and a truncated power-law electron
distribution \citep{BD97}.  The true picture is
probably some combination of these various models.  

Deep in the potential well of
Sgr~A* and pervading the central parsec of the Milky Way,
there exists a cluster of a few dozen
early-type stars \citep{SMBH90,GTKKT96},
which is dominated by the IRS~16 assemblage of probable Wolf-Rayet 
(WR) stars \citep{Najarroetal97}.
Numerous observations and models 
\citep{HKS82,AHH90,GKBW91,YM92,Najarroetal97,HRT98,PMMR01}
provide evidence that this cluster
is producing a composite hypersonic wind of 
$\dot M_\mathrm{w}\simgt10^{-3}\;\dot M_{\sun}$ yr$^{-1}$.

Some fraction
of $\dot M_\mathrm{w}$ will be trapped by the supermassive blackhole
located near
the middle of the stellar cluster.  However, all of the above
models for Sgr~A* require a mass accretion rate that is orders of magnitude less
than $\dot M_\mathrm{w}$.
If {\sl any} of these models are correct, then for
the lifetime of the early-type cluster's winds and in the
absence of any observed large-scale outflow from Sgr~A*,
some part of the cluster's winds
has been accumulating in the central parsec without being accreted by the blackhole.  
It is the purpose of this Letter to
investigate this gas in the light of recent X-ray observations of the GC.  


\section{The Mass Accumulation Rate}

The gas supply rate into the central few parsecs, based on
stellar wind measurements, is at least
$\dot M_\mathrm{w}\sim10^{-3}\;\dot M_{\sun}$ yr$^{-1}$. 
Sgr~A West is an HII region in the central parsec and contains
the streamers of the ``mini-spiral'', while
the ``circumnuclear disk'' (CND) is a dense, clumpy, and asymmetric ring-like 
feature which extends for more than 7 pc.  The CND has an inner radius of $\simeq 1.5$ pc 
and surrounds Sgr~A* and Sgr~A West.  Both the CND and Sgr~A West may be additional
sources of infalling gas \citep{MDZ96,VD01}.
The Bar, part of the ``mini-spiral'' and located
$\sim 0.1$ pc south of Sgr~A*, may be intercepting some fraction
of the cluster's winds.
In sum, $\dot M_0\sim10^{-3}\;\dot M_{\sun}$ yr$^{-1}$
is a conservative lower limit to the total rate which gas is being
supplied to the central parsecs.  Most of this gas, in the absence
of shocks and subsequent radiative cooling, is unbound and will
carve out a central cavity within Sgr~A West.  A
fraction of $\dot M_0$, $\dot M_\mathrm{A}$,
is trapped but not necessarily accreted by Sgr~A*.

Assuming static spherical wind sources and calculating
the fraction of each star's wind that is captured by Sgr~A*,
\citet{QNR99} estimated that $\dot M_\mathrm{A} \simgt 10^{-5} M_{\sun}$ yr$^{-1}$.
A capture radius can be defined by
\begin{equation}\label{eq:RA}
R_\mathrm{A} \equiv {{2 GM}\over{v_\mathrm{w}^2}}
\simeq 0.04\;\mathrm{pc} \simeq 10^5 R_\mathrm{s} \simeq 1\arcsec\;,
\end{equation}
and the Schwarzschild radius by
\begin{equation}\label{eq:RS}
R_\mathrm{s} \equiv 2 G M / c^2 \;,
\end{equation}
where $c$ is the speed of light and $v_\mathrm{w} (\sim700$km~s$^{-1})$ is the velocity of the supersonic 
wind flowing past a
centralized object of mass $M$ (taken to be $2.6\times10^6 M_{\sun}$).
Stellar motions, wind-wind collisions, radiative
cooling, and other unidentified wind sources may substantially alter
this estimate.
In addition, the
mass-loss rates of the cluster members may have been
overestimated by a factor of a few
\citep{Morrisetal00}.  
The rotation of the cluster stars
around Sgr~A* \citep{GPEGO00} may
result in angular momentum
support of their winds (but see below) and the strong magnetic fields in the
central parsec may also play a role in supporting the gas
against accretion.  Nonetheless, it appears that
$\dot M_\mathrm{A} \sim 10^{-5} M_{\sun}$ yr$^{-1}$ is
a good estimate of 
the minimum amount of gas that
is being trapped in the central arcsecond by the potential well of Sgr~A*.

According to all of the accretion models for Sgr~A*, the actual
accretion rate, $\dot M$, onto the blackhole is much less than
$\dot M_\mathrm{A}$.  In fact, the accretion rate through a radius
of $R_\mathrm{min} \sim $ 1 mas
is thought to be orders of magnitude smaller than the above estimate for $\dot M_\mathrm{A}$.
There is no evidence of
any outflow from Sgr~A* on scales larger than $R_\mathrm{min}$,
so even the presence of a
mini-jet or similar small-scale outflow
would not alter the fact that $\simgt 10^{-5} M_{\sun}$ yr$^{-1}$
is being trapped -- but, seemingly, not accreted -- by the gravitational
potential of the
blackhole.
If the wind sources are WR
stars as suggested by \citet{Tamblynetal96}
(but see below), with a WR wind
lifetime of
$t_\mathrm{w} \sim 10^5$ yrs \citep{MM87}, the total mass of gas that
has accumulated between $R_\mathrm{A}$ and $R_\mathrm{min}$ is
$\sim \dot M_0 t_\mathrm{w}/30 = \dot M_\mathrm{A} t_\mathrm{w} \equiv M_\mathrm{g} \simeq 3 M_{\sun}$.

Recent {\sl Chandra} X-ray observations \citep{Baganoffetal01} show that the 
average
ionized gas density $\sim 1 R_\mathrm{A}$ from Sgr~A* is only $\sim10^2$ cm$^{-3}$.
If $M_\mathrm{g}$ has accumulated
within $R_\mathrm{A}$, then the average number density of the captured
gas is $\bar n  \sim 10^4 \;
\mathrm{cm}^{-3}\;$.
The most likely explanation for this discrepancy is that $t_\mathrm{w} \ll 10^5$ yrs and
the GC is presently in a state of quiesence.


\section{The Cyclic Feeding of Sgr~A*}

A few million years ago, the GC underwent a burst of star formation 
\citep{Tamblynetal96} which resulted in a number of clusters of massive
stars (IRS~16, the Arches Cluster, and the Quintuplet Cluster).  Remnants
of this mini-starburst can still be seen in GC star-forming regions 
such as the ``50 km/s cloud''
\citep{SLA92, Mezgeretal89}.  The central early-type cluster may have formed
more than a parsec away from Sgr~A* and then, over a million years, spiraled
inward via dynamical friction \citep{G01}.  The disrupted remnant of the
parent molecular cloud may be seen as the ``ionized gas halo'' which fills
the central 10 parsecs \citep{APG99}.  

Then, $\sim10^4$ yrs ago, a 13--20$M_{\sun}$ star, located
$\sim 5$ pc east of Sgr~A*, exploded in
a mixed-morphology
type II supernova (SN), sweeping up the ``ionized gas halo'' and
producing what we now see as Sgr~A East \citep{Maedaetal00}.
The eastern edge of the explosion has been confined by the
``50 km/s cloud'' while the western edge has interacted with
Sgr~A* and the central cluster.
Indeed, a few thousand years ago, after overwhelming the winds 
from the mass-losing stars of the
early-type cluster, the dense frontal shock 
of the explosion swept over Sgr~A*, triggering a period
of high accretion rate and X-ray luminosity for the blackhole \citep{Baganoffetal01}.  

This period of high accretion lasted until the ram pressure of the
front shock near Sgr~A* dropped beneath that of the
central early-type cluster's winds ($\sim 10^{(-7)-(-6)}$ dynes cm$^{-2}$).
Based on SN models by \citet{PDFH01}, this ``binging'' by Sgr~A* lasted
$10^{2-3}$ yrs.  For the last thousand years or so, after the passage of the dense
shell, 
the IRS~16 and other early-type stars have
been ``purging'' the central parsecs of the hotter, less dense post-shock SN cavity material.
Sgr~A West is $\sim 1$ pc in radius, consistent with the volume these
stars could have cleared in $\sim10^3$ yrs while the arms of the ``mini-spiral''
are tendrils of infalling gas from this inherently unstable process.
Thus, $t_\mathrm{w} \sim 10^3$ yrs and $\bar n \sim 10^2 \mathrm{cm}^{-3}$, consistent
with the X-ray observations.  The lower density also reduces
the expected IR thermal bremsstrahlung
emission from the central arcsecond to below that which is observed \citep{MREG97}.
Note that radiation pressure from the early-type
cluster is not likely to influence the accumulation
of gas in the central arcsecond.  For radiation pressure
to be significant, the luminosity of the cluster, estimated to
be $\simlt 10^8 L_{\sun}$, would need to be
larger than the Eddington luminosity for Sgr~A*.  This is
almost certainly not the case \citep{LSGH99}.

\section{Discussion}

Since the trapped gas is not actually accreting 
onto Sgr~A*, it implies that dissipation of angular momentum
in the central arcsecond is, at present at least, inefficient.  
Over the next $10^5$
yrs, the density of the ISM in the central arcsecond will
increase as the winds from the early-type cluster fill the region.
Eventually, sufficient dissipation may result in
the formation of a more standard accretion disk, turning our Galaxy 
into a low level Seyfert.
However, if the stars in the central few parsecs formed at the
same time, it is
likely that another SN will explode in less than $10^5$ yrs, truncating
the accumulation process and triggering another ``binge and purge'' cycle.

If, as recently proposed \citep{PMMR01}, the early-type cluster stars are 
primarily Luminous Blue Variables (LBVs)
rather than WR stars, the shorter \mbox{duration}
of the LBV phase would imply that the cluster has been producing
massive winds for $\simlt 10^4$ yrs.  However, the winds from
LBVs are slower ($\sim 200$ km s$^{-1}$) and more massive 
($\sim 10^{-3} \dot M_{\sun}$ yr$^{-1}$) than WR winds, 
with a resulting value of
$\bar n$ that is still $\sim 50$ times larger than the
previous estimate which assumed the wind sources to be WR stars.  
Nonetheless, if the IRS~16 stars have entered the LBV
phase only within the last few centuries, the mass of trapped gas
would be sufficiently reduced to meet the X-ray observational limits.
If so, Sgr~A* might become significantly brighter in the next
few centuries as the LBV winds continue to fill the accretion region.
However, IRS~7, a red supergiant $\simeq 0.3$ pc from Sgr~A*, has
a tail that points away from IRS~16 and Sgr~A*.
Models \citep{YM92,DH94} 
show that a GC wind of $\sim 500$ km s$^{-1}$ is needed to produce
this tail.  Such a wind would require $\sim500$ yrs to reach
IRS~7 from IRS~16, implying that the wind sources are not exclusively
LBVs.

Another possibility is that the gas flows in the
central few arcseconds are not in equilibrium due to an explosive event $\sim 10^{3-4}$
yrs ago.  As mentioned above, Sgr~A East is a probable mixed-morphology type II
supernova remnant (SNR) that appears to envelope 
Sgr~A West, the compact HII region which contains both Sgr~A* and the early-type
cluster \citep{Gossetal89}.  When the progenitor of Sgr~A East, located
$\sim$ 5 pc from Sgr~A*, exploded $\sim 10^4$ yrs
ago, the strong frontal shock could have cleared the central parsec
of its accumulation of wind as it swept by Sgr~A* $\sim10^{3}$ yrs ago
\citep{Maedaetal01}.  With a temperature of 2~keV and a density
of $\sim 10$ cm$^{-3}$ \citep{Maedaetal01}, the pressure of the 
rarified gas in the cavity of the SNR is $\simlt 10^{-7}$ dyne cm$^{-2}$
while the ram pressure due to the wind from
an individual cluster member at a distance of 
$R_\mathrm{A}$ is on average $\simgt 10^{-6}$ dyne cm$^{-2}$.  Thus, after
the passage of the dense frontal shock, the cluster wind would have
overwhelmed the gas in the SNR cavity and refilled
the central $R_\mathrm{A}$ over the last $\sim 10^{3}$ yrs.
Sgr B2, an X-ray reflection nebula near the GC, seems to have been illuminated
by Sgr~A* $\sim 500$ yrs ago, suggesting that the frontal shock passed by
Sgr~A* at that time, dramatically increasing the emission of the blackhole
\citep{SMP93,Koyamaetal96,MKSTM00}.
The frontal shock was likely to be slow and dense enough to result in
Sgr~A* trapping enough gas to be accreting
at near its Eddington rate for $\sim 10^3$ yrs {\citep{Maedaetal01,Baganoffetal01}.

Since IRS 7 is presently interacting with the wind from IRS~16, 
the central $R_\mathrm{A}$ is likely to be filled with IRS~16 winds.
Thus, the rarified cavity
wind from Sgr~A East is unlikely to be the present source of accretion
for Sgr~A* as suggested by \citet{Baganoffetal01}.
However, if IRS 7 is interacting with a wind from Sgr~A* rather than
IRS~16, it would be evidence of recent activity near the blackhole,
suggesting that the IRS~16 winds have been reasserting themselves
only in the last few centuries, if at all.
A continual outflow from Sgr~A* rather than IRS~16
would clear the central arcsecond 
of gas but would require
a fast and/or dense
outflow on scales greater than $R_\mathrm{min}$;
there is no observational evidence of such a wind.

In any case, IRS~16C, a WR star only a few arcseconds from Sgr~A*, is alone producing
a sufficient wind to accumulate almost $10^{-5} M_{\sun}$ yr$^{-1}$ within $R_\mathrm{A}$.
Assuming the blackhole is accreting only a fraction of this, the upper limit
on the time the wind from this star has been accumulating in the
central arcsecond is $10^4$ yrs.  Since there are dozens of young massive stars
with heavy winds in the central parsec, it is likely that $t_\mathrm{w} \simlt 10^3$ yrs.

It is thought that Sgr~A West and Sgr~A* lie only slightly
in front of Sgr~A East since the 90 cm emission from the expanding shell of Sgr~A East
is seen in absorption at the location of Sgr~A West \citep{yusefzadehetal99}.
However, a north-south tongue of emission which cuts through Sgr~A West
and approaches Sgr~A*, can be seen.   If the proposed scenario is correct, this
tongue corresponds to the leading edge of the SNR and thus should expand
over the next few decades as the shell moves more completely in front of Sgr~A West.

It has been assumed that the early-type cluster is not 
displaced from Sgr~A* by more than a few arcseconds
along the line of sight.  Proper motion
and radial velocity observations \citep{GPEGO00} support this and
in fact suggest the orbits of the cluster stars are nearly Keplerian.
If cluster members are on Keplerian orbits around Sgr~A*, their
orbital velocities
might be an appreciable fraction of their wind velocities.
This may add sufficient angular momentum to the winds
so that a larger fraction of the gas
never reaches the capture radius.  However, the density of 
ionized gas falls off outside $\sim 1\arcsec$ \citep{Baganoffetal01}, suggesting that only a fraction of
$\dot M_\mathrm{A}$ is being trapped outside the capture radius due to support from angular momentum
or magnetic fields.

\section{Conclusions}

A fraction of the winds from the few dozen early-type stars in the central parsec of
the GC is trapped by the potential well of Sgr~A* but is not currently being accreted by the
blackhole.  Unlike an isolated cluster,
this gas has insufficient ram pressure or thermal pressure to escape so that
gas has been accumulating in 
the central $R_\mathrm{A} \sim 1\arcsec$ of the Galaxy.
Independent of the details of the flow, the estimated density of this
gas is inconsistent with IR and X-ray observations unless the accumulation
has been proceeding for $\simlt 10^3$ yrs.
This implies the gas flows in the central $\sim 1\arcsec$ are not
in equilibrium and the present low mass accretion rate onto Sgr~A*
is a short-lived phase.

Due to the tendency to form more high mass stars when deep in
a potential well, as well as the early-type stars from the present mini-starburst,
the SN rate in the GC is thought to be $\simgt 10^{-5}$ yr$^{-1}$ \citep{TR93}.
In addition, starbursts may recur in the GC every $\sim 10^{7-8}$ yrs 
\citep{O94,MGB99}.
If so, within each cycle of star formation, the sub-cycle of ``binge and purge''
may repeat on a time-scale 
of $\simlt 10^{5}$ yrs.  This episodic behaviour should 
appear in any galaxy with
star formation near a central supermassive blackhole.  Similar feedback mechanisms
have been proposed for elliptical galaxies \citep{CO01}.

In the case of the GC, one would then expect that 1) the 90 cm emission near Sgr~A* to visibily
brighten over the next few decades as the shell of the SNR moves completely in front 
of Sgr~A West, 2) the emission from Sgr~A* itself should increase over the next $\simgt 
10^3$ yrs at all frequencies as the cluster's winds again accumulate, and 3) a SN
should detonate in the next $\simlt 10^5$ yrs, leading to a brief ($\sim 10^3$ yrs)
period of high accretion and the start of the next ``binge and purge'' cycle.

\begin{acknowledgements}
This work was supported by PPARC and has made use of
NASA's Astrophysics Data System Abstract Service.  
I would like to thank J.M. Pittard for {\sl many} useful discussions
and S. Markoff and B.A. Cohen for comments on the manuscript.
\end{acknowledgements}

\bibliographystyle{apj}
\bibliography{where}

\end{document}